\documentclass[11pt]{article}
\input epsf
\usepackage{cite}
\usepackage[dvips]{epsfig}
\newcommand{\beq}{\begin{equation}}   
\newcommand{\eeq}{\end{equation}}      
\newcommand{\beqn}{\begin{eqnarray}}
\newcommand{\eeqn}{\end{eqnarray}}
\newcommand{\mbf}[1]{\mbox{\boldmath $#1$}}
\newcommand{\ba}{\bar{k}_1}
\newcommand{\bb}{\bar{k}_2}
\newcommand{\bh}{\bar{h}}
\newcommand{\C}{\cal C}
\newcommand{\dif}{{\rm d}}
\newcommand{\Ga}{\Gamma}
\newcommand{\intR}{\int_{-\infty}^{\infty}}
\newcommand{\hyp}{\,{}_2\hspace{-1pt}F_1{}}
\newcommand{\ka}{\kappa}
\newcommand{\kb}{\bar{\kappa}}
\newcommand{\kk}{\mbf{k}}
\newcommand{\la}{\lambda}

\newcommand{\N}{\cal N}
\newcommand{\R}{\cal R}
\newcommand{\rhb}{\bar{\rho}}
\newcommand{\rr}{\mbf{r}}

\begin{document}
DESY 01-XXX                          \hfill ISSN XXXX--XXXX \\

\vspace*{0.5 cm}
\begin{center}   
\begin{Large}   
\vspace*{0.5cm}  
{\bf Diffractive $\eta_c$ photo-and electroproduction
with the perturbative QCD Odderon.}
\\   
\end{Large}   
\vspace{0.5cm}   
J. Bartels$^a$ \footnote   
        {Supported by the TMR Network "QCD and Deep Structure of Elementary 
     Particles"},  
M.A.Braun$^b$ \footnote   
        {Supported by the NATO grant PST.CLG.976799} ,
D.Colferai$^a$
and G.P.Vacca$^a$ \footnote{Supported by the Alexander von Humboldt Stiftung,
by the TMR Network "QCD and Deep Structure of Elementary Particles" and by the 
NATO grant
PST.CLG.976799}\\
$^a$ II. Inst. f. Theoretische Physik,    
Univ. Hamburg, Luruper Chaussee 149, D-22761 Hamburg\\  
$^b$ St.Petersburg University,
     Petrodvoretz, Ulyanovskaya 1, 198504, Russia.
\end{center}   
\vspace*{0.75cm} 
\begin{abstract}
\noindent
Using a set of new odderon states, we calculate
their contribution to the diffractive $\eta_c$ photo- and electroproduction
process. Compared to previous simple 3-gluon exchange calculations
we find an enhancement of about one order of magnitude
in the cross section. The $t$-dependence of the cross section exhibits
a dip structure in the small $t$ region.
\end{abstract}

\section{Introduction}
The existence of the Odderon ~\cite{Lukaszuk:1973nt},
the partner of the Pomeron which is odd under charge conjugation $C$, is an
important prediction of perturbative QCD. In the leading order, the Odderon
appears as a bound state of three reggeized gluons.
Its experimental observation is a strong challenge for the experimentalists.
Promising scattering processes where the exchange of the Odderon
may be seen include the difference between the $p-p$ and $p-\bar p$
cross-sections and the diffractive production of particles with a C-odd
exchange, such as photo- and electroproduction of pseudoscalar mesons (PS).
From the theoretical point of view, some of the latter processes are of
particular interest, namely those where the presence of a large momentum
scale provides some justifications for the use of perturbative QCD.
This includes, in particular, the diffractive production of charmed
pseudoscalar mesons, for example the $\eta_c$.
Correspondingly, a large amount of literature has been devoted
to this class of diffractive processes. For large photon
virtualities $Q^2$, for heavy mass PS mesons (such as $\eta_c$), and for
large momentum transfers the relevant impact factors for the
transition $\gamma (\gamma^*) \to$PS have been be calculated perturbatively
~\cite{Czyzewski:1997bv}.
As to the Odderon exchange, first studies have used the simplest
form, the exchange of three noninteracting gluons in a C=-1 state.
Another line of Odderon investigations, persued by the Heidelberg group,
uses a non-perturbative model for the Odderon, based on the idea of
a ``stochastic QCD vacuum'' ~\cite{heidelberg}.
Numerical estimates for the cross sections
turn out to be very different in these two approaches: the nonperturbative
Odderon models tend to give substantially larger cross sections
~\cite{heidelberg,Czyzewski:1997bv,Engel:1998cg}.
Evidently, to guide the experimental search for signatures of the Odderon
we have to clarify these discrepancies.
In this paper we follow the perturbative approach and make use of the
recently discoverd new Odderon solution.
 
The perturbative QCD Odderon has a rather long history.
After several variational studies ~\cite{Gauron,Armesto:1997xz}
a first analytic solution to the
Odderon equation was constructed by Janik and Wosiek ~\cite{Janik:1999xj}
and verified by Braun {\it et al} \cite{Braun:1999mg}. It belongs to the
lowest non-zero eigenvalue of the conformal integral of motion, $Q_3$.
This solution has an intercept slightly below unity and, most important,
vanishes if two of the three gluons are at the same point. This property leads
to the disappointing result that
this solution cannot couple to the perturbative $\gamma \to$PS vertex and
therefore is irrelevant for photo and electroproduction of PS mesons.
It may, however, play its role in purely hadronic processes, such as
$pp$ or $p\bar{p}$ scattering.
Recently a new solution for a bound state of three reggeized gluons with the
Odderon quantum numbers has been found ~\cite{Bartels:2000yt},
which is quite different from the previous one.
It corresponds to $Q_3=0$, has intercept unity, and, most
important, it does couple to the $\gamma \to$PS transition vertex.
The structure of the wave function of this solution is rather peculiar,
so that one may expect substantial changes compared to the exchange of three
noninteracting gluons. The purpose of this paper is to perform an analytic
and numerical study of the exchange of this new odderon solution for the
diffractive $\eta_c$ photo- and electroproduction. We follow
the approach of ~\cite{Czyzewski:1997bv}, by replacing the three
noninteracting gluons by the new Odderon state.
As the main results, our cross sections are an order of magnitude
larger than those of ~\cite{Czyzewski:1997bv,Engel:1998cg}; also,
the $t$-dependence of our cross section exhibits an interesting dip
structure which is not present in the case of noninteracting gluons. 
\section{The perturbative QCD Odderon which couples to $\gamma$-PS transition.}
Let us first briefly recapitulate the main properties of the new odderon 
solution. For details we refer to ~\cite{Bartels:2000yt}. We remind that
this solution has been shown ~\cite{Vacca:2000bk} to be a particular case 
of a more general class of solutions: in the large $N_c$-limit there exist 
relations between eigenstates of different number of reggeized gluons, 
which connect solutions of different symmetry properties, and the Odderon 
solution is the simplest case that satisfies these relations.

In momentum space the Odderon wave function is constructed from the
known Pomeron solutions $E^{(\nu,n)}$ \cite{BFKL},
which have the eigenvalues
\beq
\chi(\nu,n)= 
\bar{\alpha_s}\left( 2\psi(1)-\psi(\frac{1+|n|}{2}+i\nu)
-\psi(\frac{1+|n|}{2}-i\nu) \right),\ \
\bar{\alpha}_s=\frac{N_c\alpha_s}{\pi}.
\label{eigen}
\eeq
One proves that
\beq
\Psi^{(\nu,n)}(\mbf{k}_1,\mbf{k}_2,\mbf{k}_3)=c(\nu,n)
\sum_{(123)} \frac{(\mbf{k}_1+\mbf{k}_2)^2}{\mbf{k}_1^2\mbf{k}_2^2} 
E^{(\nu,n)}(\mbf{k}_1+\mbf{k}_2,\mbf{k}_3), 
\label{oddwave}
\eeq 
indeed satisfies the Odderon equation and has the same intercept
(\ref{eigen}), provided the Pomeron wave function is
odd under the interchange of its arguments.
This restricts the values of $n$ to odd numbers. The 
normalization factor $c$ in (\ref{oddwave})  can be chosen in such a way that
the Odderon wave function will have the same norm as the Pomeron function
\beq
\langle \tilde{\Psi}^{(\nu,n)}|\Psi^{(\nu',n')}\rangle=
\langle \tilde{E}^{(\nu,n)}|E^{(\nu',n')}\rangle=w(\nu,n)
\delta(\nu-\nu')\delta_{nn'},
\label{scalarp}
\eeq
where the $w(\nu,n)$ are known ~\cite{Lip}.
In (\ref{scalarp}) the scalar product is defined as the integral  
of the two wave functions in momentum space, where the bra-vector has to be 
amputated (marked by a tilda). Condition (\ref{scalarp}) leads to
\beq
c(\nu,n)=\frac{1}{(2\pi)^{3/2}}
\sqrt{\frac{g_s^2 N_c}{-3\chi(\nu,n)}},\ \ 
\label{cnorm}
\eeq
When including in (\ref{oddwave}) the colour structure $d_{abc}$
one should change the normalization by a factor
$\sqrt{N_c}/\sqrt{(N_c^2-4)(N_c^2-1)}$.

In order to construct the full Greeen function we clearly need to know the
complete set of solutions of the Odderon equation.
At the moment we only have the symmetric solutions of
~\cite{Janik:1999xj} and the new solutions (\ref{oddwave}).
The former ones are orthogonal to the photon 
impact factor and so irrelevant for our problem.
At present we do not know if any other solution exists,
apart from (\ref{oddwave}),
which couples to the $\gamma-PS$ transition vertex. So our results are
strictly speaking restricted to the contribution of the exchange of the
Odderon states with the wave function (ref{oddwave}).
Normalizing the Green function 
to reduce in the small coupling limit ($\alpha_s\to 0$) to 
\[\frac{1}{\mbf{k}_1^2\mbf{k}_2^2\mbf{k}_3^2}
\delta(\mbf{k}_1-\mbf{k}'_1)\delta(\mbf{k}_2-\mbf{k}'_2).\]
and having in mind (\ref{scalarp}) we find the part of the
Green function corresponding to (\ref{oddwave}) in the form:
\[
G_3(y|\mbf{k}_1,\mbf{k}_2,\mbf{k}_3|\mbf{k}'_1,\mbf{k}'_2,\mbf{k}'_3)=\]
\beq
\sum_{{\rm odd}\ n}\int_{-\infty}^{+\infty} d\nu e^{y\, \chi(\nu,n)}
\frac{(2\pi)^2(\nu^2+n^2/4)}{[\nu^2+(n-1)^2/4][\nu^2+(n+1)^2/4]}
\Psi^{(\nu,n)}(\mbf{k}_1,\mbf{k}_2,\mbf{k}_3)
{\Psi^{(\nu,n)}}^*(\mbf{k}'_1,\mbf{k}'_2,\mbf{k}'_3) \, .
\label{greenf}
\eeq
\section{The BFKL function in the momentum space.}
In this section we present the BFKL Pomeron eigenstates
in the momentum representation, which we need in order to construct the 
Green's function in (\ref{greenf}). This wave function
is well known in the coordinate space ~\cite{Lcft}
where its form is dictated by conformal invariance
\beq
E^{(h,\bar{h})}(\mbf{r}_{10},\mbf{r}_{20})=
\left(\frac{r_{12}}{r_{10}r_{20}}\right)^h
\left(\frac{\bar{r}_{12}}{\bar{r}_{10}\bar{r}_{20}}\right)^{\bar{h}},
\label{pom_coord}
\eeq
where $\mbf{r}_{10}=\mbf{r}_1-\mbf{r}_0$ etc,
$h=(1+n)/2+i\nu$,
$\bar{h}=(1-n)/2+i\nu$, and the standard complex notation for the 
two-dimensional vector is used on the right-hand side.
Fourier transforming to momentum space one finds (see Appendix 1)
\beq
\tilde{E}_{h\bh}(\kk_1,\kk_2)= \tilde{E}_{h\bh}^{A}(\kk_1,\kk_2)+
 \tilde{E}_{h\bh}^{\delta}(\kk_1,\kk_2),
\label{pom_mom}
\eeq
where the first term denotes the analytic contribution, and the second one
stands for the $\delta$-function terms.
The analytic part is given by
\beq
\tilde{E}_{h\bh}^{A}(\mbf{k}_1,\mbf{k}_2)=C\Big(X(\mbf{k}_1,\mbf{k}_2)+
(-1)^nX(\mbf{k}_2,\mbf{k}_1)\Big),
\label{pom_an}
\eeq
where $h=(1+n)/2+i\nu$ and $\bh=(1-n)/2+i\nu$ are the conformal weights.
The coefficient $C$ is given by
\beq
C=\frac{(-i)^n}{(4\pi)^2}h\bh (1-h)(1-\bh)\Gamma(1-h)\Gamma(1-\bh).
\label{fullC}
\eeq
The expression for $X$ in complex notation is given in terms of 
the hypergeometric functions
\beq
X(\mbf{k}_1,\mbf{k}_2)=\left(\frac{k_1}{2}\right)^{\bh-2}
\left(\frac{\bb}{2}\right)^{h-2}F\left(1-h,2-h;2;-\frac{\ba}{\bb}\right)
F\left(1-\bh,2-\bh;2;-\frac{k_2}{k_1}\right).
\label{fullX}
\eeq
The $\delta$-function part is
\beq
\tilde{E}^{\delta}_{h\bh}(\mbf{k}_1,\mbf{k}_2)=
\Bigl[ \delta^{(2)}(\mbf{k}_1) +(-1)^n \delta^{(2)}(\mbf{k}_2) \Bigr]
\frac{i^n}{2\pi} 2^{1-h-\bh} \frac{\Gamma(1-\bh)}{\Gamma(h)} 
q^{\bh-1} q^{*\, h-1 } \, , \quad \mbf{q}= \mbf{k}_1+\mbf{k}_2 . 
\label{pom_delta}
\eeq
Let us note that, when the Pomeron couples to an impact factor of a
colourless object, $\delta$-function terms in the pomeron wave function do not
give any contribution. However in our calculations it turns out that these
terms play an important role.

In order to calculate the couplings of the exchanged Odderon to the proton 
and to the  $\gamma \to\eta_c$ impact factors we need to know the
Pomeron wave function $E^{(\nu,n)}$ in momentum space for $n=\pm 1$ and
around the value $\nu=0$. 
We start with the analytic part; we need to look at
(\ref{fullX}) which, for $n=1$, reads
\beq
X(\mbf{k}_1,\mbf{k}_2)=\left(\frac{k_1}{2}\right)^{i\nu-2}
\left(\frac{\bb}{2}\right)^{i\nu-1}
F\left(-i\nu,1-i\nu;2;-\frac{\ba}{\bb}\right)
F\left(1-i\nu,2-i\nu;2;-\frac{k_2}{k_1}\right),
\eeq
and we perform a Taylor expansion around the point $\nu=0$.
The complicated expression (\ref{pom_an}) is drastically simplified 
at small values of $\nu$.  
In lowest order it is linear in $\nu$:
\beq
E_1^{A}(\mbf{k}_1,\mbf{k}_2)=\frac{\nu}{2\pi^2 q}\left(\frac{1}{k_1\bb}-
\frac{1}{k_2\ba}\right).
\label{order0}
\eeq
This function is odd in the azimuthal angle (i.e. antisymmetric under 
$\varphi \to \varphi+\pi$). So it is orthogonal
to the two impact factors which are azimuthally even.
For this reason a non-zero contribution only comes from the terms
quadratic in $\nu$. Omitting those of them which have the same structure
as (\ref{order0}) we find
\beq E_2^{A}(\mbf{k}_1,\mbf{k}_2)=
\frac{i\nu^2}{2\pi^2q}\Big[
\frac{1}{\mbf{k}_2^2}\ln \mbf{k}_1^2-\frac{1}{\mbf{k}_1^2}\ln \mbf{k}_2^2
+\left(\frac{1}{\mbf{k}_1^2}-\frac{1}{\mbf{k}_2^2}\right)\ln q
+\left(\frac{1}{\bb\mbf{k}_1^2}-\frac{1}{\ba\mbf{k}_2^2}\right)
\bar{q}\ln\bar{q}\Big].
\label{E2A}
\eeq
The $\delta$-function part is finite for $\nu \rightarrow 0$:
\beq E_0^{\delta}(\mbf{k}_1,\mbf{k}_2)=
\Bigl[ \delta^{(2)}(\mbf{k}_1) - \delta^{(2)}(\mbf{k}_2) \Bigr]
\frac{i}{2 \pi}\frac{1}{q}.
\label{Edelta}
\eeq
The constant $C$ in (\ref{fullC}) (for $n=1$) becomes: 
\beq
C= \frac{1}{(4\pi)^2}\nu(1+\nu^2)\Gamma^2(1-i\nu).
\label{appr1}
\eeq
To construct the Green's function one has also to consider
\beq
\chi(\nu,\pm1)= -\nu^2 \, 2 \bar{\alpha} \zeta(3)+ {\cal O}(\nu^4)
\label{appr2}
\eeq
and the $\nu$-dependence in the denominators of the integration measure 
in (\ref{greenf}):
\beq
[\nu^2 + (n-1)^2/4][\nu^2 + (n+1)^2/4]= \nu^2 (1+\nu^2)
\label{appr3}
\eeq

\section{The transition amplitude}
We study the process of the diffractive photo- or electroproduction
of $\eta_c$ on the proton: $\gamma^*p \to \eta_c p$.
It will be assumed that the proton remains intact, although
it would be rather easy to include also its low-lying excitations.
The differential cross-section is given by the formula
\beq
\frac{d\sigma}{dt}(\gamma(\gamma^*)+p\rightarrow\eta_c+p)=
\frac{1}{16\pi s^2}\ \frac{1}{2}\sum_{i=1}^2|A^i|^2,
\label{cross1}
\eeq
where $A^i$, $i=1,2$ is the photoproduction amplitude for a given transverse
polarization $i$ of the photon $i=1,2$. The electroproduction cross-section
can be obtained from (\ref{cross1}) in a trivial manner
(see ~\cite{Czyzewski:1997bv}).
The photoproduction amplitude $A^i$ is given by a convolution
of the two impact factors, $\Phi_p$ and $\Phi_{\gamma}^i$, for the proton 
and for the $\gamma\rightarrow\eta_c$ transition, resp., with
the Odderon Green function:
\beq
A^i=\frac{s}{32}\frac{5}{6}\frac{1}{3!}\frac{1}{(2\pi)^8}
\langle \Phi^i_{\gamma}|G_3|\Phi_p\rangle.
\label{ampli}
\eeq
We shall assume that the c.m. energy squared, $s$, is much greater than
the scales in the transition vertex: $s>>Q^2,m^2_c,t$.
Using the definition of ~\cite{Czyzewski:1997bv},
namely $x=(m^2_{\eta_c}+Q^2)/(s+Q^2)$ and $y=\log{1/x}$, we are in the 
low-$x$ limit $x \ll 1$.

The matrix element on the rhs of (\ref{ampli}) involves the 
integration over $\nu$ coming from the Green function (\ref{greenf}).
This integration can be done  in the saddle point 
approximation, since $y>>1$.  
The leading contribution will obviously come from the smallest values of 
$|n|$, i.e. $n=\pm1$ 
and small $\nu$, when the integral acquires a form
\beq
\int d\nu e^{-\nu^2 \beta \, y}I(\nu),\ \ \beta=2\alpha_s\zeta(3)
\eeq 
and $I(\nu)$ denotes the nonexponential part of the integrand.
Since we expect the dominant contribution to the cross section to come
from the kinematical region where $t$ is not large, 
there is one more 
large momentum scale in the problem, apart from the energy
$\sqrt{s}$, namely,
$M=\sqrt{Q^2+4m_c^2}$, which provides some basis for the use of perturbation 
theory.
In principle, the position of the saddle point $\nu_s$ depends upon the
relation between these two large scales.
However in our kinematical region of small $x$, $M$ is much
smaller than $\sqrt{s}$, so that $\nu_s$ is close to zero.
As a result, we have to calculate the matrix elements of the two impact 
factors with the Odderon wave function with $|n|=1$ and small $\nu$.

The impact factor corresponding to the transition $\gamma\rightarrow
\eta_c$ can be calculated perturbatively. Referring the reader to the
original paper ~\cite{Czyzewski:1997bv} for the details,
we only quote the final result:
\beq
\Phi_{\gamma}^i=b \, \epsilon_{ij} \frac{q_j}{\mbf{q}^2}
\left( \sum_{(123)} 
\frac{(\mbf{k}_1 + \mbf{k}_2 - \mbf{k}_3) \cdot \mbf{q}}{Q^2+4m_c^2 + 
(\mbf{k}_1 + \mbf{k}_2 - \mbf{k}_3)^2}
-\frac{\mbf{q}^2}{Q^2+4m_c^2 + \mbf{q}^2} \right) \, .
\label{impactgamma}
\eeq 
with $\mbf{q}=\mbf{k}_1 + \mbf{k}_2 + \mbf{k}_3$ and
\beq
b=\frac{16}{\pi}e_cg_s^3\frac{1}{2}m_{\eta_c} b_0.
\eeq
Here $e_c=(2/3)e$ is the electric charge of the charmed quark, and
$g_s$ is the strong coupling constant. The constant $b_0$ 
can be determined from the known radiative width
 $\Gamma(\eta_c \to \gamma\gamma)$= 7 KeV:
\beq
b_0=\frac{16\pi^3}{3e_c^2}\sqrt{\frac{\pi\Gamma}{m_{\eta_c}}}.
\eeq
Following ~\cite{Bartels:2000yt} we denote
\beq
\varphi(\mbf{k},\mbf{k}') = 
\frac{(\mbf{k}-\mbf{k}')(\mbf{k}+\mbf{k}')}
{Q^2+4 m_c^2 +(\mbf{k}-\mbf{k}')^2}. 
\eeq
Using the form (\ref{oddwave}) of the Odderon wave function one can show that
the convolution of the Odderon wave function with the impact factor for the
transition $\gamma \to \eta_c$ reduces to a
convolution of the function $\varphi$ and the Pomeron function $E$
~\cite{Bartels:2000yt}:
\beqn 
\langle \Phi_{\gamma \to \eta_c}^i |\Psi^{(\nu,n)} \rangle &&=
b\epsilon_{ij} \frac{q_j}{\mbf{q}^2}
\frac{1}{c(\nu,n)} \int d^2 \mbf{k} \,
\varphi(\mbf{k},\mbf{q}-\mbf{k}) E^{(\nu,n)} (\mbf{k},\mbf{q}-\mbf{k}) 
\nonumber \\
&&\equiv - b\epsilon_{ij} \frac{q_j}{\mbf{q}^2} \frac{1}{c}
\frac{1}{M} (|t|)^{i \nu} \frac{i}{\pi}
V^{(\nu,n)}_{\gamma}\left({|t| \over M^2}\right),
\label{oddphot}
\eeqn
where the Pomeron function is given in (\ref{pom_mom}), and 
we have rescaled the momenta in the integrand by $q=\sqrt{|t|}$.

Note that (\ref{oddphot}) is sensitive to the $\delta$-function term
present in the Pomeron wave function.  
In fact in ~\cite{Bartels:2000yt} 
(\ref{oddphot}) has been proven by  explicitly using the Pomeron
wave function in the coordinate representation, which does contain
such terms.

We could have chosen to calculate 
$\langle \Phi_{\gamma \to \eta_c}^i |\Psi^{(\nu,n)} \rangle$ 
in a different way, directly from (\ref{oddwave}): 
\beq
\langle \Phi_{\gamma \to \eta_c}^i |\Psi^{(\nu,n)} \rangle=
3 c\, \int d^2\mbf{l} \, 
E^{*\,(\nu,n)}(\mbf{l},\mbf{q}-\mbf{l})g(\mbf{l}),
\label{oddphotbis}
\eeq
where 
function $g(\mbf{l})$ is defined by
\beq
g(\mbf{l})= \int d^2\mbf{k}\, \frac{\mbf{l}^2}{\mbf{k}^2(\mbf{l}-\mbf{k})^2}
\Phi_{\gamma}^i(\mbf{k},\mbf{l}-\mbf{k},\mbf{q}-\mbf{l}) \, ,
\label{fun2phot}
\eeq
factor $3$ comes from the symmetry of the impact factor and
$\mbf{k}_1=\mbf{k}$, $\mbf{k}_2=\mbf{l}-\mbf{k}$ and $\mbf{k}_3=
\mbf{q}-\mbf{l}$.
  
Now the $\delta$-function terms in $E$ give no contribution.
Indeed, $g(\mbf{0})=0$ because of the $\mbf{l}^2$ factor present in 
(\ref{fun2phot}), and
$g(\mbf{q})=0$ as the impact factor vanishes for $\mbf{k}_3=\mbf{0}$.
As a result, in the calculation of (\ref{oddphotbis}), one may freely 
add or subtract such $\delta$-function pieces in the Pomeron wave function 
~\cite{Mueller-Tang}.
At $\nu=\pm 1$ and small $\nu$  factor $c$ behavies
as $1/\nu$ (see (\ref{cnorm}) and (\ref{appr2}) ). Therefore, to find
the $\gamma -PS$ form-factor in the lowest (first) order in $\nu$ one
needs to know the BFKL function $E^{(\nu,,\pm 1)}$ in the second order if
one uses (\ref{oddphotbis}), but only in the zeroth order if one uses
(\ref{oddphot}). One can show that both ways lead to the same answer.
However, (\ref{oddphotbis}) needs numerical computation, and
(\ref{oddphot}) gives the result in a trivial manner due to the simple
structure of the zeroth order BFKL function (\ref{Edelta}).
At $|n|=1$ and $\nu\to 0$ we find
\beq
V_{\gamma}^{(0,\pm 1)}\left(\frac{t}{M^2}\right) = 
\frac{\sqrt{|t|/M^2}}{1+|t|/M^2} \eeq
In Fig. 1 we show a plot of this function in the region 
$0 \le |t|/M^2 \le 10$.
We have verified by numerical computation that,
starting from (\ref{oddphotbis}) and using for
$E^{* \, (\nu,n)}$ only the analytic part, $E_2^A$, we get exactly
the same answer for $\nu \rightarrow 0$.

The proton impact factor is non-perturbative. We use the parametrization 
proposed in ~\cite{Czyzewski:1997bv}:
\beq
\Phi_p=d\Big[F(\mbf{q},0,0)-\sum_{i=1}^3F(\mbf{k}_i,
\mbf{q}-\mbf{k}_i,0)+2F(\mbf{k}_1,\mbf{k}_2,\mbf{k}_3)\Big],
\label{proton_if}
\eeq
with
\beq
F(\mbf{k}_1,\mbf{k}_2,\mbf{k}_3)=
\frac{2a^2}{2a^2+(\mbf{k}_1-\mbf{k}_2)^2+
(\mbf{k}_2-\mbf{k}_3)^2+(\mbf{k}_3-\mbf{k}_1)^2},
\label{Fp}
\eeq
$d=8(2\pi)^2\bar{g}^3$,
and the scale parameter $a= m_{\rho}/2$.
From the comparison with the two gluon exchange model for hadronic 
cross-sections the authors of ~\cite{Czyzewski:1997bv} estimate
$\bar{g}^2/(4\pi)=1$.
The impact factor (\ref{proton_if}) satisfies the basic requirement
that it vanishes when any of the three gluon momenta goes to zero.
The calculation of the scalar product of the Odderon wave function with the 
proton impact factor is more cumbersome, since it
amounts to the integration over the three-gluon phase space. 
Using the symmetry
of the proton impact factor and the explicit form of the Odderon
wave function (\ref{oddwave}) one finds
\beq
 \langle \Phi_p|\Psi^{(\nu,n)}\rangle=
3 c\, d\int d^2\mbf{l} \, 
E^{*\,(\nu,n)}(\mbf{l},\mbf{q}-\mbf{l})f(\mbf{l})=  c \, d 
\frac{1}{(2a^2)^{1/2}} (t)^{-i \nu} \frac{(i \nu^2)^*}{2\pi^2}
V^{(\nu,n)}_p\left({|t| \over 2a^2}\right),
\label{oddprot}
\eeq
where
\beq
f(\mbf{l})= \int d^2\mbf{k}\, \frac{\mbf{l}^2}{\mbf{k}^2(\mbf{l}-\mbf{k})^2}
\big[F(\mbf{q},0,0)-\sum_{j=1}^3F(\mbf{k}_j,\mbf{q}-\mbf{k}_j,0)
+2F(\mbf{k}_1,\mbf{k}_2,\mbf{k}_3)\Big]
\label{funprot}
\eeq
and, as before, $\mbf{k}_1=\mbf{k}$, $\mbf{k}_2=\mbf{l}-\mbf{k}$ and 
$\mbf{k}_3=
\mbf{q}-\mbf{l}$. To obtain the last expression in (\ref{oddprot}) we have 
again rescaled the momenta in the integral with respect to $q$ 
and extracted the factor $i \nu^2 /2\pi^2$ having in mind that at small $\nu$
the contribution to (\ref{oddprot}) starts with the second order term of
the BFKL function (\ref{E2A}). 
Note that the  integral (\ref{funprot}) is infrared finite, since the
square bracket vanishes if any of the gluon momenta go to zero. However,
individual terms inside the square bracket are infrared divergent.
Clearly, as in (\ref{oddphotbis}), only the analytic part of the Pomeron 
function $E^A$ contributes to (\ref{oddprot}).

\unitlength=1mm
\begin{figure}
\label{Fig1}
\centering
\begin{picture}(80,50)
\put(0,0){\includegraphics[width=8cm]{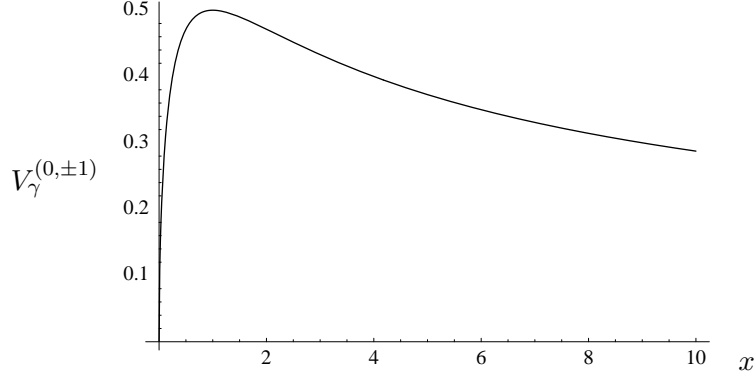}}
\put(-15,24){{ $V_{\gamma}^{(0,\pm 1)}$ }}
\put(83,0){$x$}
\end{picture}
\caption{Numerical results for the coupling of the Odderon to the 
$\gamma^* \to \eta_c$ impact factor (defined in (\ref{oddphot}), 
as a function of the scaled variable $x=|t|/(Q^2+4 m_c^2)$.}
\end{figure}
     
Let us consider the rhs of (\ref{funprot}).
Two of the five terms are simple,
since the $F$'s do not depend on the integration variable. They  give
\beq
f_1=
\Big(F(\mbf{q},0,0)-F(\mbf{q}-\mbf{l},\mbf{l},0)\Big)
\int d^2\mbf{k}\frac{\mbf{l}^2}{\mbf{k}^2(\mbf{l}-\mbf{k})^2}=
\Big(F(\mbf{q},0,0)-F(\mbf{q}-\mbf{l},\mbf{l},0)\Big)2\pi
\ln\frac{\mbf{l}^2}{m^2}.
\label{f1}
\eeq
The two remaining terms in the sum inside the square brackets in 
(\ref{funprot}) give identical contributions. Their sum is given by
\beq
f_2= 
-\frac{2}{3}a^2\int d^2\mbf{k}\frac{\mbf{l}^2}{\mbf{k}^2(\mbf{l}-\mbf{k})^2}
\frac{1}{(\mbf{k}-\mbf{q}/2)^2+\epsilon},
\label{f2}
\eeq
where $\epsilon=\frac{1}{12}\mbf{q}^2+\frac{1}{3}a^2$.
The last term in the sum gives:
\beq
f_3= 
\frac{2}{3}a^2\int d^2\mbf{k}\frac{\mbf{l}^2}{\mbf{k}^2(\mbf{l}-\mbf{k})^2}
\frac{1}{(\mbf{k}-\mbf{l}/2)^2+\delta}.
\label{f3}
\eeq
where $\delta=\mbf{l}^2-\mbf{l}\mbf{q}+(\mbf{q}^2+a^2)/3$.
Both integrals, (\ref{f2}) and (\ref{f3}), thus reduce to a generic one
\beq
I=\int d^2\mbf{k}\frac{1}{(\mbf{k}^2+m^2)[(\mbf{k}-\mbf{l})^2+m^2]
[(\mbf{k}-\mbf{p})^2+b^2]},
\label{triangle}
\eeq
corresponding to a general two-dimensional triangle diagram.
We have introduce here a mass $m$ as an infrared regulator.
Using the standard Feynman parametrization this integral can be
transformed into an one-dimensional one which, after separating the 
infrared divergent contributions, can be done numerically. 
Some details about this procedure are discussed in Appendix 2
(note that (\ref{f3}) actually corresponds to a collinear
case which, in principle, can be calculated analytically).

The resulting function $f(\mbf{l})=f_1+f_2+f_3$ is used to calculate the 
integral (\ref{oddprot}).
Here only the analytic part, $E_2^A$, contributes.
The calculation has been done numerically.
We present the results for $V_p^{(0,\pm 1)}$ in Fig. 2. 
One observes that $V_p$ changes sign at $|t| \approx 0.07$ GeV$^2$. As a 
consequence, the cross section will vanish at this point. Of course, 
this property is literally true only in the limit of asymptotically large 
energies where one can neglect the contributions of all other states
$\nu \neq 0$ and $|n|>1$.

\begin{figure}
\label{Fig2}
\centering
\begin{picture}(80,50)
\put(0,0){\includegraphics[width=8cm]{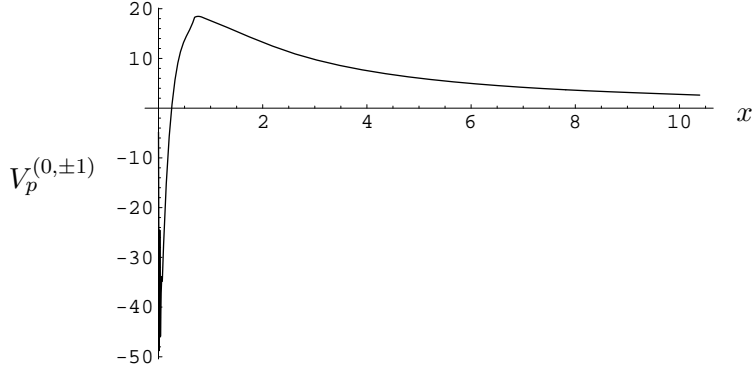}}
\put(-15,24){{ $V_p^{(0,\pm 1)}$ }}
\put(83,33){$x$}
\end{picture}
\caption{Numerical results for the coupling of the Odderon to the proton,
defined in (\ref{oddprot}), as a function of the scaled variable $x=|t|/2a^2$.}
\end{figure}

\section{Numerical Results and Discussion}
To find the final cross-sections from (\ref{cross1}) and 
(\ref{ampli}), we do the saddle point integration 
over $\nu$,  then 
take the square module of the amplitude (\ref{ampli})
and do the sum over the polarizations in (\ref{cross1}). The latter step 
provides a factor $1/|t|$ (from the prefactors in (\ref{impactgamma})).
The normalization factors $c(\nu,n)$
of the Odderon solution which are contained in the Green's function
(\ref{greenf}) cancel when, in (\ref{oddphot}), the scalar product of the
Odderon wave  function and the photon impact factor is computed. 
Collecting all pieces of our cross section formula we find
\beq
\frac{d\sigma}{dt}(\gamma(\gamma^*)+p\rightarrow\eta_c+p)=
\frac{2^4 \cdot 5^2 }{3^7}\frac{1}{(2\pi)^8}
\frac{\alpha_{em} \alpha_s^2 b_0^2}{\zeta(3)y}\frac{m_{\eta_c}^2}{(Q^2+4 m_c^2)2 a^2}
\frac{1}{|t|}|V^{(0,\pm 1)}_\gamma(t)|^2|V^{(0,\pm 1)}_p(t)|^2.
\eeq
The differential cross sections for the two cases $Q^2=0$  and 25 
GeV$^2$ and $\sqrt{s}\approx 300$ GeV are shown in Fig.3. 
We have taken $\alpha_s$ at the scale $m_c^2 +Q^2$
(in ~\cite{Czyzewski:1997bv} at $Q^2=0$ the scale was $m_c^2$).
Independently of $Q^2$ the cross sections show a dip at small $|t| 
\approx 
0.07$ GeV  and a maximum at 
$|t| \approx 0.22$ GeV (roughly of the
order of $(m_{\rho}/2)^2$. The dip comes from the 
zero present in the Odderon-proton coupling (Fig. 2). Its origin 
seems to be related to the symmetry properties of the Odderon solution:
as it can be seen from (\ref{oddwave}), the Odderon wave function is a sum of
three terms, each of which contains an antisymmetric Pomeron eigenfunction.
A similar structure is present in the photon impact factor (\ref{impactgamma}),
whereas the proton impact factor (\ref{proton_if}) is completely symmetric.
The convolution of the Odderon wave
function with the photon impact factor has no zero in $t$, whereas the
convolution with the proton leads to such a zero. Whether this feature is an
artifact of the simple model for the proton impact factor that we have used,
or whether it represents a general property of the Odderon-proton coupling  
we do not know. In our calculation, this dip is present in the leading
high energy approximation; it may be that at finite energies 
(e.g. at HERA) the dip is (partially) filled by the exchange of 
nonleading Odderon states.   

At $t=0$ the cross section vanishes, as in
the case of a simple three gluon exchange. As one can see from 
Fig. 3, this happens at quite small $t$  and cannot be seen in the figure.
A more detailed analysis of the behaviour in the region of very small $t$
requires, probably, a more accurate evaluation of the $\nu$-integral in the
3-gluon Green's function.
It would, however, not affect too much the dip structure or the value
of the integrated cross section.\\

\begin{figure}[h]
\label{Fig3}
\centering
\begin{picture}(80,50)
\put(0,0){\includegraphics[width=8cm]{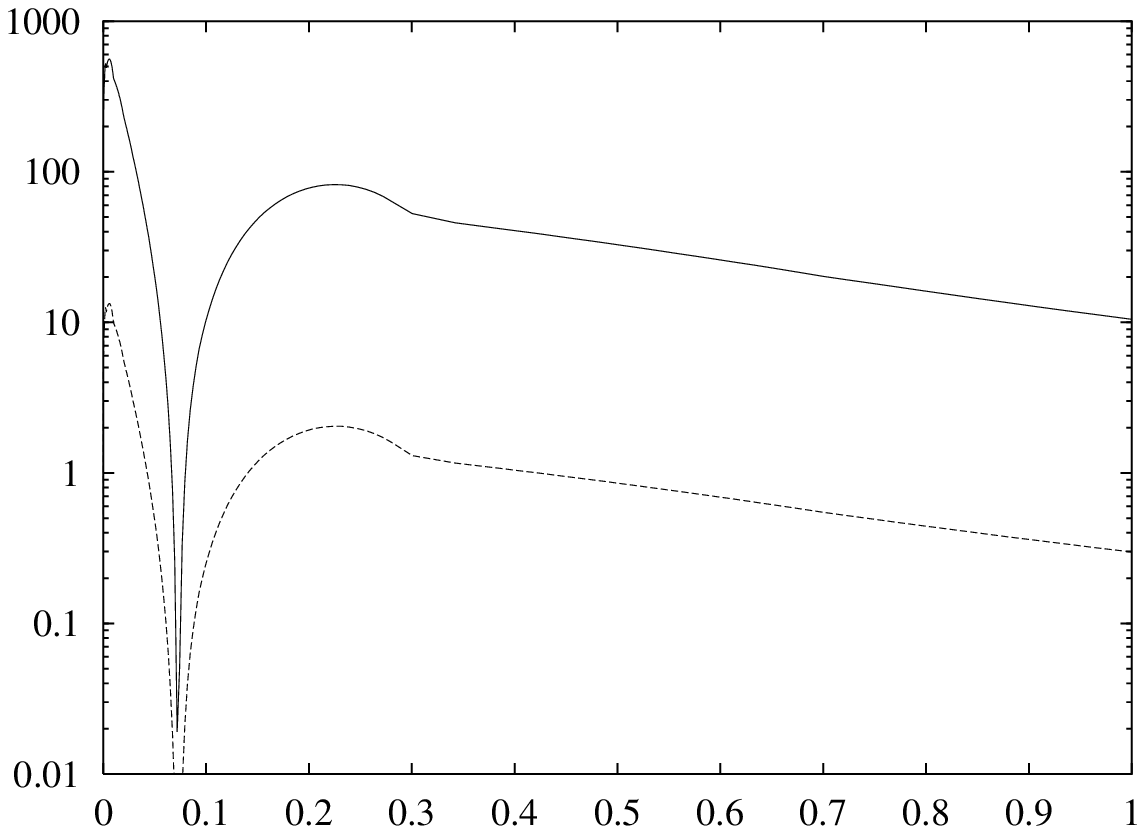}}
\put(-5,30){$ \displaystyle{d\sigma \over dt}$ }
\put(80,0){$|t|$}
\end{picture}
\caption{The differential cross sections (in pb $/$ GeV$^2$). 
The upper curve refers to $Q^2=0$.}
\end{figure}

For the integrated cross sections we find $50$ pb and $1.3$ pb at
$Q^2=0$ and  25 GeV$^2$, respectively.
Compared to the value of 11 pb  predicted 
for $Q^2=0$ with a simple three-gluon exchange ~\cite{Czyzewski:1997bv}
 we find an enhancement of about 5 times.
For $Q^2=25$ GeV$^2$ the differential 
 cross sections in ~\cite{Czyzewski:1997bv} seem to
indicate that one can simply scale the $Q^2=0$ cross section by 0.01 and
obtain $0.1$pb. This implies that our cross-section is an order of 
magnitude larger.
Note that  compared to the simple
three gluon exchange,  we have a (weak) logarithmic suppression with
energy. So the obtained enhancement effect is totally due to the coupling 
of our Odderon wave function to the impact factors.

In conclusion, by comparing the exchange of three noninteracting gluons 
with the exchange of the new odderon solution we find that the interaction 
between the exchanged gluons leads to a significant change in the scattering
cross section. However, despite this improvement in accuracy, our 
numerical estimate of the $\eta_c$ cross section still suffers from 
a few theoretical uncertainties, in particular due to the Odderon-proton 
coupling. We believe that both the structure of the vertex and its overall 
normalization should be checked more 
carefully. As a possible strategy, one might study the exchange of the
Odderon in $pp$ and $p\bar{p}$ scattering at large $t$ where the use of 
perturbative QCD can be justified. Using the same model  
~\cite{Czyzewski:1997bv}, a comparison with experimental data on the 
difference of $pp$ and $p\bar{p}$ scattering fixes the  
overall normalization of the Odderon proton coupling. As to the general 
momentum structure of the vertex, the most sensitive test is the 
$t$-dependence in the small-$t$ region: the presence of a dip 
would support both the structure of the Odderon-proton vertex and 
of the Odderon state used in our calculation.\\ \\ 
\noindent
{\bf Acknowledgements:} One of us (M.B.) thanks the University of Hamburg, 
the II.Institut f\"ur Theoretische Physik, and the DESY Theory Division 
for financial support and for their kind hospitality.
\section*{Appendix 1. A momentum space representation of the Pomeron wave 
function in the non-forward direction}

The Fourier transform of the function $E^{(h,\bh)}$ given by
Eq. (\ref{pom_coord}) is defined by
\begin{equation}
\label{defTF1}
 \tilde{E}_{h\bh}(\kk_1,\kk_2)=\int{\dif^2\rr_1\over(2\pi)^2}{\dif^2\rr_2
 \over(2\pi)^2}\;
 \left({r_{12}\over r_1r_2}\right)^h \left({r^*_{12}\over r^*_1r^*_2}
 \right)^{\bh}
 e^{i(\kk_1\cdot\rr_1+\kk_2\cdot\rr_2)}\;.
\end{equation}
The integrand in (\ref{defTF1}) becomes constant
as $|\mbf{r}_{1,2}| \to \infty$ and therefore 
contains terms proportional to $\delta^2(\kk_{1})$ or $\delta^2(\kk_{2})$.
Therefore one expects to find
\beq
\tilde{E}_{h\bh}(\kk_1,\kk_2)= \tilde{E}_{h\bh}^{A}(\kk_1,\kk_2)+
 \tilde{E}_{h\bh}^{\delta}(\kk_1,\kk_2),
\eeq
where the first term denotes the analytic contribution and the second
the $\delta$-like one.

We shall at first compute the analytic part of the Fourier transform.
Since we are dealing with a distribution,
it is convenient to consider a new, regularized, object

\begin{equation}
\label{defTF2}
\tilde{E}^{(reg)}_{h\bh,h_3\bh_3}(\kk_1,\kk_2)=\int{\dif^2\rr_1\over(2\pi)^2}
{\dif^2\rr_2\over(2\pi)^2}\;
{\left(r_{12} \right)^{h_3} \over \left( r_1r_2 \right)^h} 
{\left(r^*_{12} \right)^{\bh_3} \over \left( r^*_1r^*_2 \right)^{\bh} }
 e^{i(\kk_1\cdot\rr_1+\kk_2\cdot\rr_2)}\;,
\end{equation}

with an independent conformal weigth $h_3$ for the $r_{12}$-terms.
The integral (\ref{defTF2}) 
is well defined for $\Re(h+\bh)< 2$, $\Re(h_3+\bh_3)>-2$ and
$\Re(h+\bh-h_3-\bh_3)> 0$. Strictly speaking, the last inequality holds, for 
example, for
the $\kk_2 \ne \mbf{0}$ case. When $\kk_2 = \mbf{0}$ one needs 
$\Re(h+\bh-h_3-\bh_3)> 2$.
In the calculation of the
scalar product of the Pomeron function with test functions which vanishe at the points
$\kk_1=0$ or $\kk_2=0$ 
we shall, therefore, consider the following prescription:
$\langle \Phi | E_{h\bh} \rangle= \lim_{h_3\to h} \langle \Phi | 
E_{h\bh,h_3\bh_3} \rangle=
\langle \Phi | \lim_{h_3\to h} E_{h\bh,h_3\bh_3} \rangle $.
This means that we shall be able to extract the analytic part of 
(\ref{defTF1}). 

Introducing, for both external gluons of the Pomeron wave function, the 
complex variables
\[
 r=r_x+i r_y\equiv x+i y\ \   \ka={k_x-i k_y\over2}\]
 \begin{equation} \label{compvar}
 r^*=r_x-i r_y\equiv x-i y\ \ \ka^*={k_x+i k_y\over2},
\end{equation}
and remembering that $h-\bh=n$ and $h_3-\bh_3=n_3$, we get a
double integral in
the complex plane:
\begin{equation}\label{compint}
  \tilde{E}^{(reg)}_{h\bh,h_3\bh_3}(\kk_1,\kk_2)=
  \int{\dif^{\C}\rr_1\dif^{\C}\rr_2\over(2\pi)^4}\;
 |r_1^2|^{-h} |r_2^2|^{-h} |r_{12}^2|^{h_3}  r_1^{*\,n} r_2^{*\,n}r_{12}^
 {*\,-n_3}
 e^{i(\ka_1 r_1+\ka_1^* r_1^*+ \ka_2 r_2+\ka_2^* r_2^*)}\;.
\end{equation}
By using the integral representation
\begin{equation}\label{repower}
 x^{-u}={1\over\Ga(u)}\int_0^\infty\dif\alpha\;\alpha^{u-1}e^{-\alpha x}
 \qquad(\Re(x)>0\;,\;\Re(u)>0)\;
\end{equation}
we arrive at the representation
\[
 E^{(reg)}_{h\bh}(\ka_1,\ka_2)={1\over\Ga^2(h)\Ga(-h_3)}
 \int{\dif^{\C}\rr_1\dif^{\C}\rr_2\over(2\pi)^4}\;(r_1^* r_2^*)^n
 r_{12}^{*\,-n_3}
 e^{i(\ka_1 r_1+\ka_1^* r_1^*+ \ka_2 r_2+\ka_2^*r_2^*)}\]
 \begin{equation}\label{rapEpow}
 \int_0^\infty\dif\alpha_1\dif\alpha_2\dif\alpha_3\;
 (\alpha_1\alpha_2)^{h-1}\alpha_3^{-h_3-1}
 e^{-(\alpha_1 r_1 r_1^*+\alpha_2 r_2 r_2^*+\alpha_3 r_{12} r_{12}^*)}\;,
\end{equation}
where, in order to fulfill the restrictions in Eq.~(\ref{repower}), we need
to require
\begin{equation}\label{condh}
 \Re(h)>0\;,\;\Re(h_3)<0\quad\iff\quad n>-1\;,\;n_3<-1\;.
\end{equation}

The next step is to perform a suitable change of variables in order to
do the spatial integrations. Thanks to the holomorphic separability of the
integrand,  the
natural choice would be to use $r_j$ and $r_j^*$ as independent integration
variables. This can practically be achieved in the following way: first of all
note that the
whole integrand, regarded as a function of $(x_1,y_1,x_2,y_2)$, is
analytic. Therefore, we can analytically continue to complex values of,
say, $y_1$
and $y_2$, and rotate the respective integration paths by negative angles.
This can
be done by putting $y=e^{-i\theta}w:w\in\R$ for both $j=1,2$,
and by 
simultaneously rotating the $\alpha$-integrations in such a way that
$\alpha=e^{i\phi}\beta:\beta\in\R$.
The generic 
exponential in the second line in~(\ref{rapEpow}) becomes
\begin{equation}\label{trasfexp}
 e^{-\alpha r r^*}=e^{-\alpha(x^2+y^2)}=e^{-\beta e^{i\phi}(x^2+
 e^{-2i\theta}w^2)}=
 e^{-\beta x^2 e^{i\phi}}e^{-\beta w^2 e^{i(\phi-2 \theta)}},
\end{equation}
which does not grow for large $x$ and $w$ provided
$|\phi|\leq\pi/2 \;,\;|\phi-2\theta|\leq\pi/2$. Choosing the extreme case
$\phi=\theta=\pi/2$
we set $y=-iw\;,\;\alpha=i\beta$ and perform the change of variables
\begin{equation}\label{changevar}
 \rho_j=x_j+w_j\;,\;\rhb_j=x_j-w_j\qquad(j=1,2)\;.
\end{equation}
A further remark concerns the values of the complex momenta $k_j$:
the exponential in the first line
in ~(\ref{rapEpow}) contains two factors
\begin{equation}\label{trasfmom}
 e^{i(k_x x+k_y y)}=e^{i(k_x x-i k_y w)}=e^{i k_x x}e^{k_y w}\;.
\end{equation}
In order to have a meaningful integral, the values of the
$y$-components of the momenta must be imaginary. Hence we have to introduce a 
slightly
different notation with respect to Eq.~(\ref{compvar})
\begin{equation}\label{compvarkbar}
 \ka={k_x-i k_y\over2}\quad;\quad\kb={k_x+i k_y\over2}\neq\ka^*\quad;
 \quad\ka,\kb\in\R
\end{equation}
and to consider $\ka$ and $\kb$ as independent variables.
After this replacement we get
\[
  E^{(reg)}_{h\bh}(\ka_1,\ka_2;\kb_1,\kb_2)={(-i/2)^2(2\pi)^{-4}
  \over\Ga^2(h)\Ga(-h_3)}
 \intR\dif\rhb_1\dif\rhb_2\;
 (\rhb_1\rhb_2)^n\rhb_{12}^{\;-n_3} e^{i(\kb_1\rhb_1+\kb_2\rhb_2)}\times\]
 \begin{equation}\label{Ebarvar} 
 i^{2h-h_3}\int_0^\infty\dif\beta_1\dif\beta_2\dif\beta_3\;
 (\beta_1\beta_2)^{h-1}\beta_3^{-h_3-1}\intR\dif\rho_1\dif\rho_2\;
 e^{-i(\beta_1\rho_1\rhb_1+\beta_2\rho_2\rhb_2+\beta_3\rho_{12}\rhb_{12}
-\ka_1 \rho_1-\ka_2\rho_2)}\;.
\end{equation}
The last integral in the above expression provides two delta functions
constraining
\[
\left( \matrix{\ka_1\cr\ka_2\cr}\right)=
 \left(\matrix{\beta_1+\beta_3&-\beta_3\cr
               -\beta_3&\beta_2+\beta_3\cr}\right)\!\!
 \left(\matrix{\rhb_1\cr\rhb_2\cr}\right)\iff
 \left(\matrix{\rhb_1\cr\rhb_2\cr}\right)
 ={1\over\beta_1\beta_2+\beta_2\beta_3+\beta_3\beta_1}
 \left(\matrix{\beta_2+\beta_3&\beta_3\cr
               \beta_3&\beta_1+\beta_3\cr}\right)\!\!
 \left(\matrix{\ka_1\cr\ka_2\cr}\right).
\]
If, for simplicity, we restrict ourselves to positive values of the 
$\ka$'s, it is apparent from the
second equality in the above expression that the positivity of the
$\beta$'s forces the
$\rhb$-integrals to contribute only in the positive real half-plane.
At this point we
split the $\rhb$-integral into two pieces: the first takes into account
the region
$\rhb_2<\rhb_1$ and the second the remaining one $\rhb_2>\rhb_1$.
The first contribution reads
\begin{equation}\label{first}
 \int_0^\infty\dif\rhb_1\int_0^{\rhb_1}\dif\rhb_2\;(\rhb_1\rhb_2)^n
 \rhb_{12}^{\;-n_3}
 e^{i(\kb_1\rhb_1+\kb_2\rhb_2)}\times\end{equation}\[
 \qquad\int_0^\infty\dif\beta_1\dif\beta_2\dif\beta_3\;
 (\beta_1\beta_2)^{h-1}\beta_3^{-h_3-1}
 \delta(\beta_1\rhb_1+\beta_3\rhb_{12}-\ka_1)
 \delta(\beta_2\rhb_2-\beta_3\rhb_{12}-\ka_2).
\]
We can factorize the above expression in the product of two independent
integrals by
means of the change of variables $\la_i=\beta_i \rhb_i:i=1,2,3\;,\;
(\rhb_3\equiv\rhb_{12})$,
which yields
 \[ \int_0^\infty\dif\rhb_1\int_0^{\rhb_1}\dif\rhb_2\;(\rhb_1\rhb_2)^{n-h}
 \rhb_{12}^{\;-n_3+h_3}e^{i(\kb_1\rhb_1+\kb_2\rhb_2)}\times \]
 \begin{equation}\label{intfact}
 \qquad\int_0^\infty\dif\la_1\dif\la_2\dif\la_3\;(\la_1\la_2)^{h-1}
 \la_3^{-h_3-1}
 \delta(\la_1+\la_3-\ka_1)\delta(\la_2-\la_3-\ka_2)\;.
\end{equation}
The $\rhb$-integrals are easily evaluated by setting $x=\rhb_2/\rhb_1$,
which casts
the inner integral into an integral representation of the confluent
hypergeometric function $_1\!F_1$ (see, e.g., Eq.~(13.2.1) of \cite{AbSt}).
The outer integral then becomes 
a Laplace transform of the $_1\!F_1$ times a power (see
Eq.~(7.621.4) of \cite{GrRh}). The result for the first factor of
Eq.~(\ref{intfact}) is (remember $n-h=-\bh$)
\begin{equation}\label{one}
 -i^{\bh_3-2\bh}{\Ga(1-\bh)\Ga(1+\bh_3)\Ga(2-2\bh+\bh_3)\over
 \Ga(2-\bh+\bh_3)}
 \kb_1^{\;2\bh-\bh_3-2}\hyp(1-\bh,2-2\bh+\bh_3;2-\bh+\bh_3;
 -{\kb_2\over\kb_1})\;.
\end{equation}
All the restrictions in the above formulas are automatically fulfilled
because of Eq.~(\ref{condh}).
The $\la$-integral in Eq.~(\ref{intfact}) transforms into
\begin{equation}\label{intlambda}
 \ka_1^{h-h_3-1}\ka_2^{h-1}\int_0^1\dif y\;y^{-h_3-1}(1-y)^{h-1}
 \big(1+{\ka_1\over\ka_2}y\big)^{h-1},
\end{equation}
which is just an integral representation for the hypergeometric
function (see, e.g.,
Eq.~(15.3.1) in \cite{AbSt}) and yields
\begin{equation}\label{two}
 {\Ga(h)\Ga(-h_3)\over\Ga(h-h_3)}\ka_1^{h-h_3-1}\ka_2^{h-1}
 \hyp(-h_3,1-h;h-h_3;-{\ka_1\over\ka_2})\;,
\end{equation}
provided Eq.~(\ref{condh}) holds.
The second contribution to (\ref{Ebarvar}), coming from the region
$\rhb_2>\rhb_1$, is
simply evaluated by the replacements
$\rhb_1\leftrightarrow\rhb_2\;,\;\beta_1\leftrightarrow\beta_2$,
 which give
\begin{equation}\label{second}
 \int_{\rhb_2>\rhb_1}=(-1)^{n_3}\int_{\rhb_2<\rhb_1}
 \left(\matrix{\ka_1\leftrightarrow\ka_2\cr\kb_1\leftrightarrow\kb_2\cr}
 \right)\;,
\end{equation}
where the parity factor stems from the change of sign of $\rhb_{12}$ in
the power with exponent $-n_3$.

To derive the final
expression for
the Pomeron wave function in momentum space
we have to analytically continue in the conformal
weights
to their physical values $h_3=h,\bh_3=\bh$ and
$\ka_{y}\in \R\iff$  $\kb = \ka^*$.
To do this in Eq.~(\ref{two}) 
we use the relation (see Eq.~(15.1.2) in \cite{AbSt})
\begin{equation}\label{limit}
 \lim_{h_3\to h}{1\over\Ga(h-h_3)}\hyp(-h_3,1-h;h-h_3;-{\ka_1\over\ka_2})=
 h(1-h){\ka_1\over\ka_2}\hyp(1-h,2-h;2;-{\ka_1\over\ka_2})\;.
\end{equation}
Putting together
Eqs.~(\ref{one},\ref{two},\ref{second},\ref{limit}), and
rearranging
 some $\Ga$-function factors according to the relation
\begin{equation}\label{Gammarel}
 \Ga(\bh)\Ga(1-\bh)=(-1)^n\Ga(h)\Ga(1-h)\qquad(h-\bh=n\in\N)\;,
\end{equation}
we obtain the final expression
\[
  E_{h\bh}^{A}(\ka_1,\ka_2)={h(1-h)\Ga(1-h)\bh(1-\bh)\Ga(1-\bh)
  \over i^n\,(4\pi)^2}\times
\]
\begin{equation}\label{Emom}
 \left[\ka_1^{*\,\bh-2}\ka_2^{h-2}\hyp(1-h,2-h;2;-{\ka_1\over\ka_2})
 \hyp(1-\bh,2-\bh;2;-{\ka_2^*\over\ka_1^*})+(-1)^n\{1\leftrightarrow2\}
 \right]\;.
\end{equation}

As a check, we show in the following that the above function is an
eigenfunction of the
Casimir operator of the M\"obius group.
In the coordinate representation, in complex notation, one has
\beq
\left( (r_{12})^2 \partial_1 \partial_2  +h(h-1) \right)
\left({r_{12}\over r_1r_2} \right)^h  =0,
\eeq
together with a similar equation in the antiholomorphic variables.
In the momentum representation, these equations read
are given by
\beq
\left( (\partial_{\ka_1} -\partial_{\ka_2})^2 \ka_1 \ka_2 +h(h-1)
\right) E_{h\bh}(\ka_1,\ka_2)=0,
\label{pass1}
\eeq
and an analogous result holds for its antiholomorphic counterpart.
From the general property of scaling invariance we know that 
\beq
E_{h\bh}(\ka_1,\ka_2) = \ka_1^{h-2} \ka_1^{*\,\bh-2}
E_{h\bh}(1,\frac{\ka_2}{\ka_1}),
\eeq
which is satisfied by (\ref{Emom}). Changing the variables
$\ka_1 \to p_1 \, , \, \ka_2/\ka_1 \to p_2$
transforms  (\ref{pass1}) into
\beq
\left(
\left( \partial_{p_1}-\frac{1+p_2}{p_1}\partial_{p_2} \right)^2 p_1^2 p_2
+h(h-1) \right) p_1^{h-2} E_{h\bh}(1,p_2) =0.
\label{pass2}
\eeq
Taking the derivative with respect to $p_1$, we are
left with a differential equation in the $p_2$
variable only. We represent it in terms of a new variable $y=-p_2$:
\beq
\left( y(1-y) \partial_{y}^2 +(2 - 2(2-h) y)\partial_{y}-(h-1)(h-2) \right)
E_{h\bh}(1,-y) =0.
\label{pass3}
\eeq
This is the well known hypergeometric equation.
A linearly
independent set of its solutions is given by the
Kummer solutions $u_1,u_4$ (formulas (2.9.1) and (2.9.13) in ~\cite{Bateman}):
\beq
u_1(y)=  \hyp(1-h,2-h;2;y) \quad , \quad
u_4(y)= (-y)^{h-2} \hyp(1-h,2-h;2;\frac{1}{y}).
\label{kummersol}
\eeq
These solutions match exactly the structure in (\ref{Emom}),
which is therefore a solution
of (\ref{pass1}).

One can arrive directly at (\ref{Emom}) by trying to construct 
a single-valued function  from the
two linearly independent solutions, and by further fixing the correct
normalization.
Note that  $\delta$-like terms 
do not appear in this approach, since we
treat the holomorphic and antiholomorphic momenta  as independent variables.
The fact that, in principle, the two sectors do not simply commute
(as one can see considering
 the two dimensional Poisson equation
with a $\delta$-like source) gives origin to the appearence
of the $\delta$-like terms.

Let us now consider the $\delta$-like contributions.
These are present when $\rho_1 \rightarrow \infty$ or
$\rho_2 \rightarrow \infty$. Summing these two contributions one obtains
\beq
\tilde{E}^{\delta}_{h\bh}(\mbf{k}_1,\mbf{k}_2)=
\Bigl[ \delta^{(2)}(\mbf{k}_1) +(-1)^n \delta^{(2)}(\mbf{k}_2) \Bigr]
\frac{i^n}{2\pi} 2^{1-h-\bh} \frac{\Gamma(1-\bh)}{\Gamma(h)} 
q^{\bh-1} q^{*\, h-1 } \, , \quad \mbf{q}= \mbf{k}_1+\mbf{k}_2. 
\eeq
\section*{Appendix 2. Integrals appearing in the
coupling to the nucleon}
In order to do the integral (\ref{triangle})
we use the Feynman parametrization
\[\frac{1}{ABC}=2\int_0^1xdx\int_0^1dy\frac{1}{D^3},\]
where
\[D=xyA+x(1-y)B+(1-x)C.\]
In our case
\[D=xy(\mbf{k}^2+m^2)+x(1-y)((\mbf{k}-\mbf{l})^2+m^2)+
(1-x)((\mbf{k}-\mbf{p})^2+b^2)=
(\mbf{k}-x(1-y)\mbf{l}-(1-x)\mbf{p})^2+R,\]
where
\beq
R=x(1-y)\mbf{l}^2+(1-x)(\mbf{p}^2+b^2)+xm^2-
(x(1-y)\mbf{l}+(1-x)\mbf{p})^2.
\eeq
Shifting the integration momentum and doing the momentum integration
we find the integral over $x$ and $y$
\beq
I=\pi\int_0^1xdx\int_0^1dy\frac{1}{R^2}.
\eeq
One of the integrations (say, of $y$) can be done analytically.
We present
\[
R(y)=\alpha+\beta y +\gamma y^2,
\]
where
\[\alpha=x(1-x)(l-p)^2+(1-x)b^2+xm^2,\ \ \beta=
x(2x-1)l^2+2x(1-x)pl,\ \ \gamma=-x^2l^2.\]
The discriminant $\Delta=4\alpha\gamma-\beta^2$ is
 negative, so that the integral over $y$ gives
\beq
J=\frac{\beta+2\gamma}{\Delta R(1)}-\frac{\beta}{\Delta R(0)}
+\frac{2\gamma}{\Delta\sqrt{-\Delta}}
\ln\frac{\beta+2\gamma-\sqrt{-\Delta}}{\beta+2\gamma+\sqrt{-\Delta}}
\frac{\beta+\sqrt{-\Delta}}{\beta-\sqrt{-\Delta}}.
\label{J}
\eeq

One finds that  at $x\rightarrow 0$ the integral $J$ behaves as $1/x$, so that 
the integration over $x$ is convergent around this point. At 
$x\rightarrow 1$ $\Delta$ is finite in the limit $m\rightarrow 0$. However,
in the limit $x\rightarrow 1$ both $R(0)$ and $R(1)$ behave as $1/(1-x)$.
So we have
a logarithmic divergence at $x=1$, regularized by finite $m$. Evidently only the
first two terms in (\ref{J}) lead to this divergence. Therefore we can safely 
put
$m=0$ in the third (logarithmic) term. In the vicinity of $x=1$ we find

\beq
J\sim J_0= \frac{1}{l^2}\left(\frac{1}{(1-x)(p^2+b^2)+m^2}+\frac{1}
{(1-x)[(l-p)^2+b^2]+m^2}\right).
\eeq
The final integration over $x$ of this term will give 
\beq
S=\int_0^1xdxJ_0=
\frac{1}{l^2}\Big[\frac{1}{p^2+b^2}\left(\ln\frac{p^2+b^2}{m^2}-1\right)
+\frac{1}{(l-p)^2+b^2}\left(\ln\frac{(l-p)^2+b^2}{m^2}-1\right)\Big].
\label{S}
\eeq
One easily checks that the singular contributions coming from $f_2$ and
$f_3$ are cancelled by the singular part of $f_1$ (eq. (\ref{f1})), so that 
the complete result is infrared finite.

To do the numerical calculation we   present $ J=J-J_0+J_0\equiv J_r+J_0$.
The integral over $x$ of the  term $J_r=J-J_0$ converges at $x=1$ and
can be  calculated 
numerically.
The integral over $x$ of  $J_0$ is given by (\ref{S}).


\begin{thebibliography}{99}

\bibitem{Lukaszuk:1973nt}
L.~Lukaszuk and B.~Nicolescu,
Lett.\ Nuovo Cim.\ {\bf 8} (1973) 405.

\bibitem{Czyzewski:1997bv}
J.~Czyzewski, J.~Kwiecinski, L.~Motyka and M.~Sadzikowski,
Phys.\ Lett.\ {\bf B398} (1997) 400 [hep-ph/9611225]; erratum Phys. Lett
                {\bf B411} (1997) 402.

\bibitem{heidelberg}
E.~R.~Berger, A.~Donnachie, H.~G.~Dosch, W.~Kilian, O.~Nachtmann and M.~Rueter,
Eur.\ Phys.\ J.\ {\bf C9} (1999) 491 [hep-ph/9901376].

A.~Schafer, L.~Mankiewicz and O.~Nachtmann,
UFTP-291-1992
{\it  In *Hamburg 1991, Proceedings, Physics at HERA, vol. 1* 243-251 and 
Frankfurt Univ. - UFTP 92-291 (92,rec.Mar.) 8 p}.

W.~Kilian and O.~Nachtmann,
Eur.\ Phys.\ J.\ {\bf C5} (1998) 317 [hep-ph/9712371].

M.~Rueter, H.~G.~Dosch and O.~Nachtmann,
Phys.\ Rev.\ D {\bf 59} (1999) 014018 [hep-ph/9806342].

\bibitem{Engel:1998cg}
R.~Engel, D.~Y.~Ivanov, R.~Kirschner and L.~Szymanowski,
Eur.\ Phys.\ J.\ {\bf C4} (1998) 93 [hep-ph/9707362].

\bibitem{Gauron}
P.~Gauron, L.~Lipatov and B.~Nicolescu,
Phys.\ Lett.\ {\bf B304} (1993) 334;
Z.\ Phys.\ {\bf C63} (1994) 253.

\bibitem{Armesto:1997xz}
N.~Armesto and M.~A.~Braun,
Z.\ Phys.\ {\bf C75} (1997) 709 [hep-ph/9603218].

\bibitem{Janik:1999xj}
R.~A.~Janik and J.~Wosiek,
Phys.\ Rev.\ Lett.\ {\bf 82} (1999) 1092 [hep-th/9802100].

\bibitem{Braun:1999mg}
M.~A.~Braun, P.~Gauron and B.~Nicolescu,
Nucl.\ Phys.\ {\bf B542} (1999) 329 [hep-ph/9809567].

\bibitem{Bartels:2000yt}
J.~Bartels, L.~N.~Lipatov and G.~P.~Vacca,
Phys.\ Lett.\ {\bf B477} (2000) 178 [hep-ph/9912423].

\bibitem{Vacca:2000bk}
G.~P.~Vacca,
Phys.\ Lett.\ {\bf B489} (2000) 337 [hep-ph/0007067].

\bibitem{Lip}
\bibitem{BFKL} E. A. Kuraev, L. N. Lipatov and V. S. Fadin,
               Sov. {\bf JETP 44} (1976) 443; \\
               {\bf ibid. 45} (1977) 199;\\
               Ya. Ya. Balitskii and L.N. Lipatov, Sov. J. Nucl. Phys.
               {\bf 28}, (1978) 822. 

\bibitem{Lcft}  L.N. Lipatov, {\it Pomeron in quantum chromodynamics}, in
                ``Perturbative QCD'', pp. 411-489, ed. A. H. Mueller,
                World Scientific, Singapore, 1989;
                Phys. Rep. {\bf 286} (1997) 131.

\bibitem{Mueller-Tang} A. H. Mueller and W. K. Tang,
                       Phys. Lett. {\bf B 284} (1992) 123.

\bibitem{AbSt} M.~Abramowitz and I.~Stegun, Handbook of Mathematical 
               functions, Dover, 1970.
\bibitem{GrRh} I.S.~Gradstein and I.M.~Ryshik, Summen-, Produkt- Und 
               Integraltafeln, Deutsch, 1981.

\bibitem{Bateman} Harry Bateman, Bateman Manuscript Project,
                  Vol I : Higher Transcendental Functions,
                  Erdelyi Editor, McGraw-Hill (1953)  

\end{thebibliography}
\end{document}